\documentclass[11pt]{article}
\usepackage{epsfig,rotating}
\usepackage{cite}

\def\Journal#1#2#3#4{{#1} {\bf #2}, #3 (#4)}

\def\PLB{{\em Phys. Lett.}  B}

\def\PRD{{\em Phys. Rev.} D}
\def\ZPC{{\em Z. Phys.} C}
\def\JPG{{\em J. Phys.} G}
\def\IJMPA{{\em Int. J. Mod. Phys} A}
\def\CPC{\em Comp. Phys. Com.}

\def\z{z_p}

\def\mj2{\protect\hat{s}}

\newcommand{\beq}{\begin{equation}}
\newcommand{\eeq}{\end{equation}}
 
\newcommand{\la}{\langle}
\newcommand{\ra}{\rangle}

\newcommand{\uq}{\mathrm{u}}
\newcommand{\dq}{\mathrm{d}}
\newcommand{\sq}{\mathrm{s}}
\newcommand{\q}{\mathrm{q}}

\begin{document}
\vspace*{10mm}

\hfill ANL-HEP-CP-99-88

\hfill September 1999   

\begin{center}  \begin{Large} \begin{bf}
Issues in Leading Particle  and
Charm Production \\ in DIS at HERA \footnote{ 
Proc. of the workshop "Monte Carlo Generators for HERA
Physics", DESY-PROC-1999-02, 
(Hamburg, Germany, 1998-1999), Ed. A.T.Doyle at al, ISSN
1435-8077, pp.309-320}  
\\ 
\end{bf}  \end{Large}
\vspace*{5mm}
\begin{large}
S.~V.~Chekanov \\  
\end{large}

\vspace*{5mm}
HEP Division, 
Argonne National Laboratory,~9700~S.~Cass~Avenue, \\ 
Argonne, IL~60439~ USA\\

\end{center}
\begin{quotation}

\vspace{1.0cm}

\noindent
{\bf Abstract:}
A Monte Carlo simulation 
based on ${\cal O}(\alpha_{\rm s})$
QCD matrix elements matched to parton showers   
shows that final-state hadrons 
in deep inelastic scattering (DIS) can be used  
to tag  events with  a single  (anti)quark recoiled  
against the proton. The method is particularly suited to study
the mean charge of leading particles, 
which is sensitive to fragmentation and sea quark contribution to 
the proton structure function. We also discuss 
methods to study the charm production in DIS using
the Breit frame. 
\end{quotation}
%
 
%%%%%%%%%%%%%%%%%%%%%%%%%%%%%%%%%%%%%%%%%%%%%%%%%%%%%%%%%%%%%%%%%%
\section{Introduction} 
%%%%%%%%%%%%%%%%%%%%%%%%%%%%%%%%%%%%%%%%%%%%%%%%%%%%%%%%%%%%%%%%%%

\begin{figure}[htb]
\vspace{1.5cm}
\begin{center}
\mbox{\epsfig{file=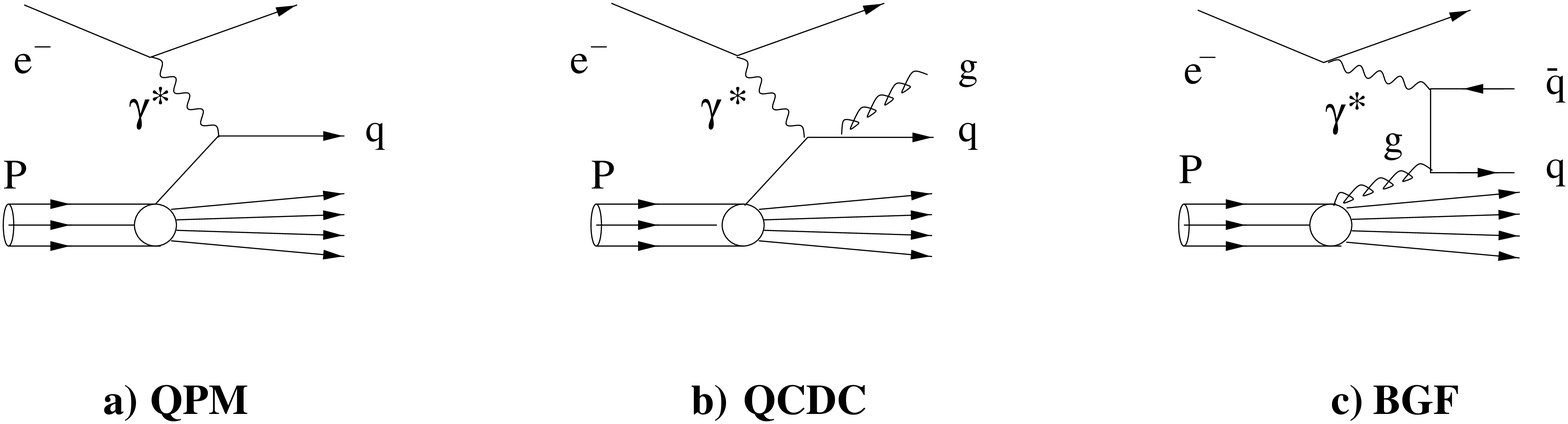,height=4.0cm}}
\end{center}
\vspace*{2mm}
\caption{{\it QPM and the ${\cal O}(\alpha_{\rm s})$ QCD processes.}}
\label{fig1}
\end{figure}

The quark-parton model (QPM) is a simple picture of
deep inelastic scattering (DIS)   
which gives a  basic framework for nucleon's parton distributions.
Taking into account quark-gluon 
interactions at ${\cal O}(\alpha_{\rm s})$ in 
the leading-order (LO) approximation, this picture is
modified by QCD Compton (QCDC) and Boson-Gluon
Fusion (BGF) processes as shown in Fig.~\ref{fig1}.
High-order perturbative QCD emissions of partons
are  usually included through the parton shower mechanism 
in the leading-log approximation. 
 
Nowadays, the hard QCD processes are well understood and can 
be modeled successfully with Monte Carlo (MC) models. 
This knowledge can be used to investigate  various  less-understood
non-perturbative effects that are often obscured  by perturbative QCD.    
For example, manifestations of instantons 
\cite{MGHERA1}, leading particles \cite{MGHERA2}
and intrinsic charm \cite{MGHERA3} in hadronic
final-states of  DIS 
are affected by the conventional perturbative QCD radiations.
 
In this paper we  discuss how to reduce  unwanted 
perturbative QCD effects related to the BGF processes in the study 
of the nature of the struck  quark  at HERA energies.
We suggest to constrain  an    
event sample  in order to obtain DIS events with  
a single struck quark recoiled  against the proton (with possible
initial- and final-state gluon radiations). After this  selection,
the resulting events  have kinematic signatures
similar to QPM/QCDC events.     
Such a selection is especially important for analysing the properties
of the struck quark which carries important  information on the proton
structure. Using this selection, we investigate the mean charge of leading 
particles, i.e. hadrons emerging from the struck quark after fragmentation. 
Being  a simple quantity to measure,  
the leading-particle charge is shown to be 
sensitive to details in  fragmentation as well as  
the sea parton distribution inside the proton. 
We also investigate the production of charm in DIS by isolating
the QPM/QCDC from the BGF source of charm production.    

The DIS  is  characterised by the
4-momentum transfer $Q^2=-q^2$ and the Bjorken
scaling variable $x=Q^2/(2 P\cdot q)$,
where $P$ is the
4-momentum of the proton.
For the QPM in the Breit frame \cite{Br}, by convention, 
the incident quark carries $Q/2$ momentum
in the positive $Z$-direction and the outgoing
struck quark carries the same momentum in the
negative $Z$-direction.
The phase space of the event can be divided into two
regions. Particles with negative $p_Z^{\mathrm{Breit}}$ components
of momenta form the current region. 
In the QPM  all these particles are produced
from  hadronisation  of
the struck quark. Particles with positive $p_Z^{\mathrm{Breit}}$  are
assigned to the target region, which is associated with the
proton remnant.

The LO QCD processes, QCDC and BGF, give  
rise to two final-state partons leading to two jets,  in 
addition to the proton remnant in  the target
region. The LO effects lead to
anti-correlations between the current- and target-region 
multiplicities \cite{chek}.

To conform the theoretical expectations discussed in this paper, 
we use LEPTO 6.5 \cite{LEP}. This model  
incorporates the  ${\cal O}(\alpha_{\rm s})$ 
LO  matrix elements matched to parton showers (MEPS).  
The parton showers  and hadronisation are described by the JETSET 7.4
\cite{jetset}. Here it is important to emphasize a few  
points: We  use  default parameters in the LEPTO, while  our 
results depend on the values of cut-offs on the LO matrix elements. 
A dependence of the results on the 
cut-offs values should be studied.
This requires a re-tuning of the LEPTO with a new set of the cut-offs.    
This has not been done yet. Secondly, a MC dependence of 
the results should also be investigated. However, the most  popular models, 
such as ARIADNE \cite{ARD} and HERWIG \cite{HEW}, 
do not contain exact matrix elements. Thus they cannot distinguish 
between QPM, QCDC and BGF on an event-by-event basis,  
which is important for our studies. In addition,  
a theoretical conclusion which might be drawn from
these models is far from clear due to  
an ambiguity in the modeling of the parton cascade. 
 
%%%%%%%%%%%%%%%%%%%%%%%%%%%%%%%%%%%%%%%%%%%%%%%%%%%%%%%%%%%%%%%%%%
\section{BGF reduction}
\label{sec:red}
%%%%%%%%%%%%%%%%%%%%%%%%%%%%%%%%%%%%%%%%%%%%%%%%%%%%%%%%%%%%%%%%%

\begin{figure}[htb]
\begin{center}
\vspace{2.5cm}
\mbox{\epsfig{file=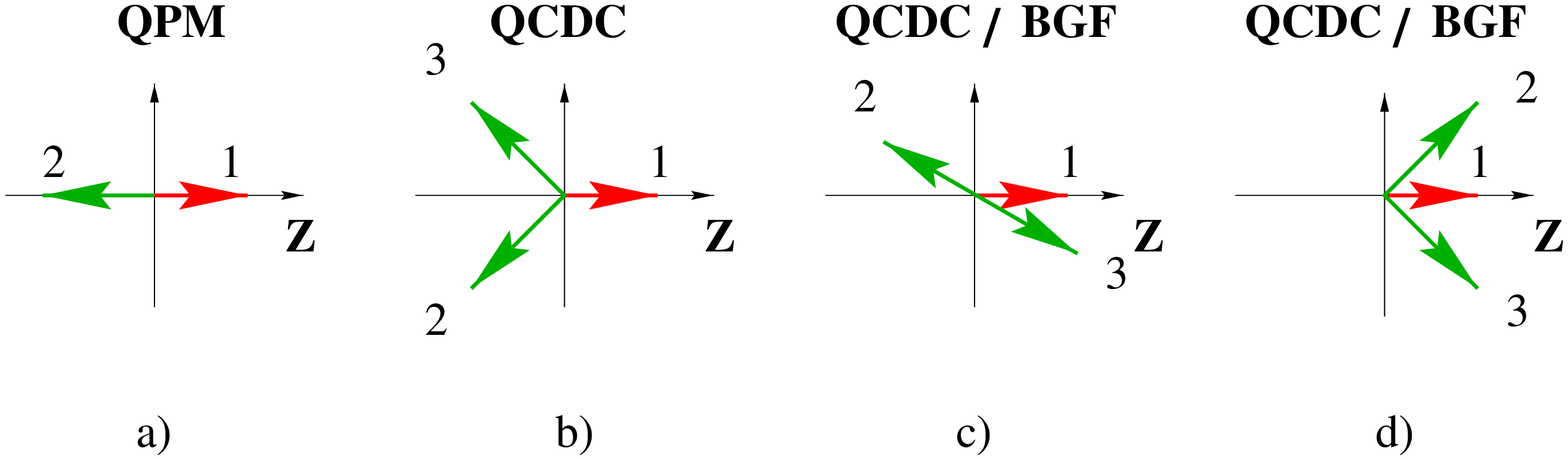,height=4.0cm}}
\end{center}
\caption{{\it Various configurations
of final-state partons at the LO in the Breit frame.
``1'' denotes the spectator partons  while
``2'' and ``3'' denote the final-state
partons from the hard QCD interactions.}}
\label{fig3}
\end{figure}

The two-parton production at the LO 
approximation can be studied in terms of five
independent variables \cite{jur}.  
Particularly important are two of them:     
$x_p=Q^2/ (2p\cdot q)$ 
and the scaling variable $\z=(p'\cdot p) / (p'\cdot q)$, 
where $p$ is the momentum of the incoming parton and $p'$
is that of the final-state parton. 
At the ${\cal O}(\alpha_{\rm s})$, 
the singularities in the two-parton  cross section
are given by \cite{jur}:
\beq
d\sigma_{2+1}^{\mbox{\protect\footnotesize BGF}}
\propto
\frac{[\z^2+(1-\z)^2][x_p^2+(1-x_p)^2]}{\z(1-\z)}, \qquad   
d\sigma_{2+1}^{\mbox{\protect\footnotesize QCDC}}
\propto
\frac{1+x_p^2\z^2}{(1-x_p) (1-\z)}. 
\eeq
In the BGF process, the poles 
are related to the emissions
of  two quarks collinear to the initial gluon 
or when the partons are soft ($\z\to 1,0$).
The QCDC diverges if the radiated gluon is collinear to initial- or
final-state quark or if it is soft ($\z$, $x_p \to 1$). 
The singularity $x_p=1$, corresponding  to the gluon  
radiated collinear to the final-state quark,  
favours the production of two partons in the current regions. 
This effect is not present 
in the BGF as illustrated in Fig.~\ref{fig3}.   
Indeed, according  to \cite{str},
let us introduce two new variables, $z_1=(1-x_p-\z )/x_p$ and
$z_2=(\z - x_p )/x_p$, which are proportional to the longitudinal
momenta of the two final-state partons in the Breit frame.  
For the QCDC singularities, these partons move to the current region 
if $z_1 < 0$ and $z_2 < 0$. 
For the singularities $\z \to 0, 1$ in the BGF cross 
section  this  situation is
impossible for  any value of $x_p$. Thus, in this limit, the BGF cannot
produce two partons in the current region. 
      
\begin{figure}[htb]
\begin{center}
\vspace{1.0cm}
\mbox{\epsfig{file=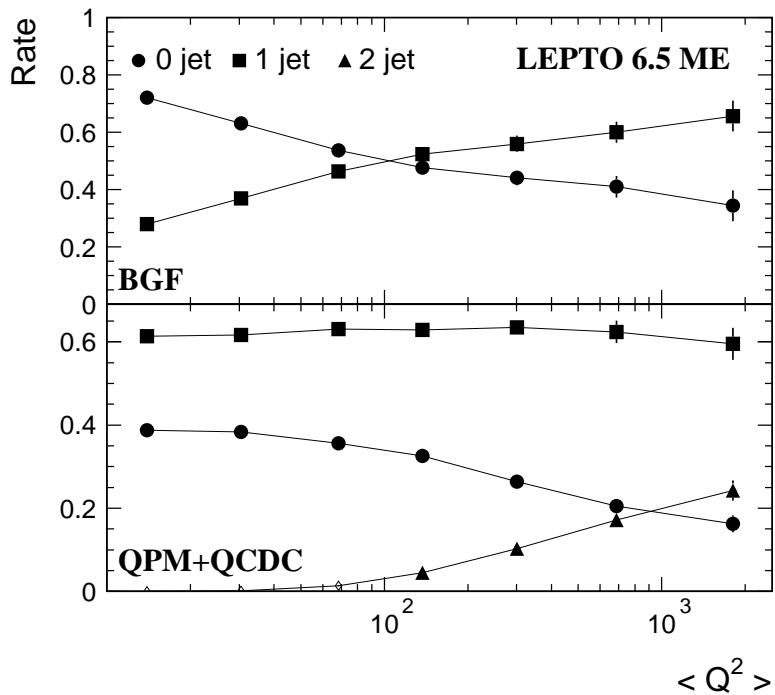,height=14.0cm}}
\end{center}
\caption{{\it The production rates of various
kinematic topologies of the LO  partons in the current
region of the Breit frame
as a function of $Q^2$. The rates are generated with the LEPTO model.
The lines are to guide the eyers.
The notation ``0 jet'' corresponds to empty current region, while
``1 jet'' and ``2 jet'' denote the production of one or two
final-state hard partons in the current region, respectively. }}
\label{fig4}
\end{figure}

To conform this  expectation, we simulated the production of 
two final-state partons  in the Breit frame using  the LO matrix elements 
implemented in  LEPTO. The parton shower and hadronisation 
were turned off. We used the GRV 
proton structure function \cite{GRV} from the PDFLIB \cite{pdf}. 
To generate  DIS events,
the energy of the positron and  that
of the proton are   chosen to be
27.5 GeV and  820 GeV, respectively.
The production rate for each parton  
configuration in the current region is shown in Fig.~\ref{fig4},  
where the notation ``jet'' 
refers to one of the LO hard partons. 
The BGF has many  events without partons 
in the current region (``0 jet''). This configuration
and that with a single  parton in the current region (``1 jet'')
are  characteristic for both BGF and QCDC. At a sufficiently high $Q^2$,
QCDC  gives rise to a fraction of   
events with  two partons (quark and gluon) 
in the current region (``2 jet'').  

\begin{figure}[htb]
\begin{center}
\vspace{1.0cm}
\mbox{\epsfig{file=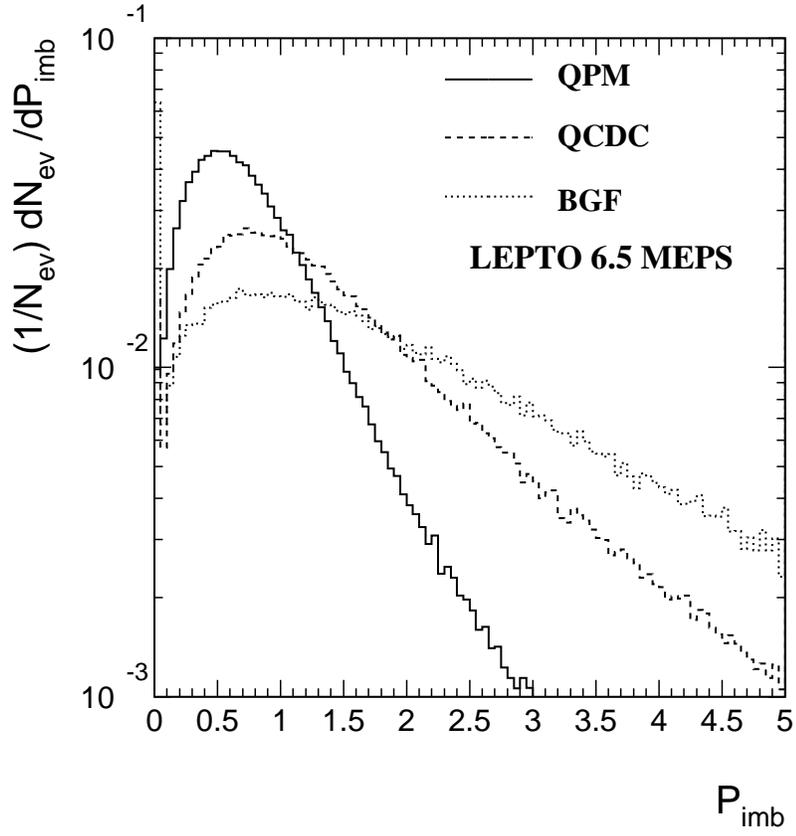,height=14.0cm}}
\end{center}
\caption{{\it The transverse imbalance of the
current-region charged particles in LEPTO MEPS for
the LO processes. The  histograms are normalised to unity.
The sharp peak at $P_{\mathrm{imb}}=0$ seen for BGF and QCDC
corresponds to events without charged  particles in the current region.}}
\label{fig5}
\end{figure}

According to this observation, the BGF
contribution to DIS can be reduced following the strategy: 

\begin{itemize}

\item 
To identify phase-space regions with  minimum QCD radiations.
DIS events at not very high $Q^2$ and a sufficiently large $x$
are the most obvious choice.     

\item
To require to detect more than one jet in the current region.     
This selection is likely to be possible at a sufficiently large $Q^2$.
At low $Q^2$, the jet structure is  less prominent.   
In this case, however,  many  BGF events have both jets in the target 
region (see Fig.~\ref{fig4}).  
Therefore, for small $Q^2$, one could simply 
require to detect  at least one or two 
final-state hadrons in the current region.

\item
To restrict the transverse momentum imbalance in  the
current region, $P_{\mathrm{imb}}=\sum_{i}p_{T,i}$, where
$p_{T,i}$ is  the  transverse momentum of the $i$th  particle in
the current region of the Breit frame  and
the sum runs over all current-region particles. 
The QPM leads to a single  jet, collinear to the $Z$-direction, while 
the QCDC can produce two jets
in the current region with well balanced transverse momenta. Therefore,  
$P_{\mathrm{imb}}\simeq 0$ for these cases. In reality, of course, the
imbalance is not zero due to high-order QCD effects, hadronisation
or resonance decays.    
Fig.~\ref{fig5} shows the current-region
transverse imbalance for  charged final-state hadrons  in
QPM, QCDC and BGF processes simulated with  LEPTO MEPS.
The BGF events  have  the broadest distribution of $P_{\mathrm{imb}}$. 
Thus, imposing a restriction on the  $P_{\mathrm{imb}}$ 
can help to reduce the BGF events in the selected subsample.    
\end{itemize}

\begin{figure}[htb]
\begin{center}
\vspace{1.0cm}
\mbox{\epsfig{file=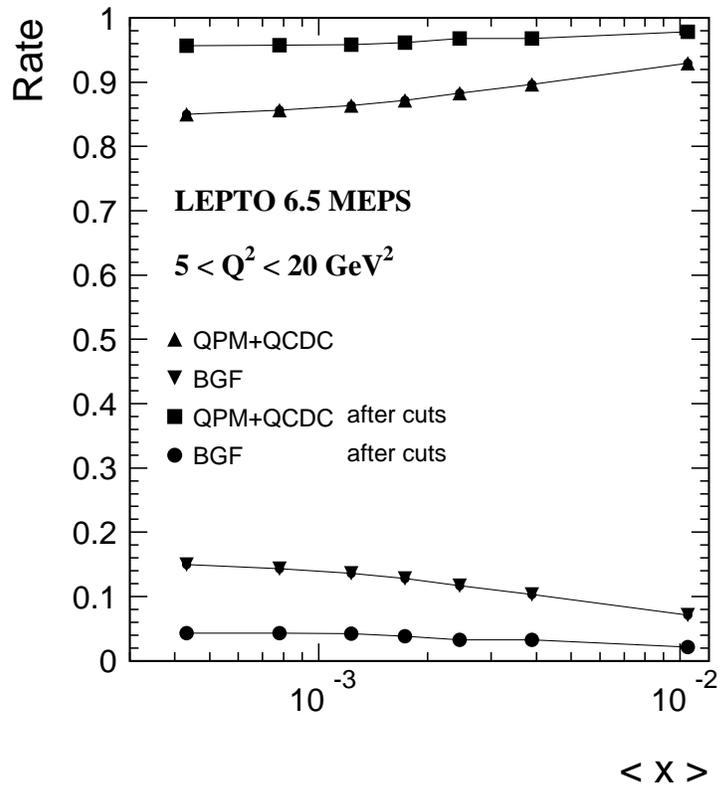,height=14.0cm}}
\end{center}
\caption{{\it The production rate of the LO QCD processes in LEPTO before
and after the selection procedure. The lines are to guide the eyers.}}
\label{fig6}
\end{figure}

Fig.~\ref{fig6} illustrates the method. Using LEPTO MEPS,
we simulated  DIS events before and after  the cuts.
We require  
to have at least two final-state charged hadrons
in the current region and $P_{\mathrm{imb}}< 0.7$, in order to keep
the transverse current-region imbalance as small as possible
but without a large reduction of the number of events
passed  this cut.   
For the default LEPTO parameters,   
the production rate of BGF varies from 
$7\%$ to $16\%$, depending on $x$.
The selection  results to  a subsample 
with only $2 - 4\%$  of the BGF events.
The efficiency of such selection is about $20\%$. 

A suppression effect exists also for the QCDC, but it is not as strong as for
the BGF. For the kinematic regions shown in  Fig.~\ref{fig6}, the
QCDC production rate varies from $4\%$ to $5\%$. After the cuts, about half
of these events survive (not shown).    

Note that this method is quite different to that in which 
one requires  to  observe  
a single jet, in addition to the remnant jet, using cluster
algorithms. Our method is intended to reduce the BGF, rather than QCDC.
In contrast, the 
requirement to observe a single jet in an event rejects both  BGF and QCDC.
In addition, our method is well suited to study DIS events 
at rather low  $Q^2$, independent of   
a jet transverse energy $E_T$ used in jet reconstruction. 

%%%%%%%%%%%%%%%%%%%%%%%%%%%%%%%%%%%%%%%%%%%%%%%%%%%%%%%%%%%%%%%%%%
\section{Mean Charge of Leading Hadrons}
%%%%%%%%%%%%%%%%%%%%%%%%%%%%%%%%%%%%%%%%%%%%%%%%%%%%%%%%%%%%%%%%%%

Below we estimate the  average  electric
charge of a  leading current-region particle, i.e. a hadron  
with a  minimum  value of $p_Z^{\mathrm{Breit}}$.   
We expect that it is this hadron that retains  main properties
of the struck quark.  
To demonstrate  a sensitivity of the leading-particle charge  
to  details in fragmentation and  sea quark production, 
we shall  obtain  this quantity   
analytically and using a MC simulation after
the BGF reduction.  

%%%%%%%%%%%%%%%%%%%%%%%%%%%%%%%%%%%%%%%%%%%%%%%%%%%%%%%%%%%%%%%%%%
\subsection{Fragmentation in the QPM} 
%%%%%%%%%%%%%%%%%%%%%%%%%%%%%%%%%%%%%%%%%%%%%%%%%%%%%%%%%%%%%%%%%%

First let us find  the mean charge of the
struck quark in the QPM assuming that the proton
consists of three valence quarks,  
$p=(\uq\uq\dq )$. We define  
the electric charge $q_i$ of  quarks $\uq$ and $\dq$ 
as $q_{\uq}=2/3$ and $q_{\dq}=-1/3$, respectively. 
The virtual boson couples to the  valence quark with a  
probability proportional to $q_i^2$. Using this fact and  that
there are two valence quarks $\uq$ and only one quark $\dq$,
the probabilities $P_{\q}$ for the virtual boson to interact 
with the valence quark $\q=\uq, \dq$  are   
$P_{\uq} = 8/9$ and  $P_{\dq}= 1/9$, respectively.   
The mean charge of the current region is 
\beq
\la q \ra = \sum_{\uq , \dq} P_i q_i = 5/9 \simeq 0.55.  
\label{c2}
\eeq
After the hard interaction, the  struck quark $\uq$  couples  to
an antiquark $\bar{\q}_i=\bar{\dq}$, $\bar{\uq}$, $\bar{\sq}$ 
(the contribution from heavy quarks is neglected).  
Taking into account the suppression factor $0.3$ for strange quarks, 
similar to \cite{jetset},   
the probabilities $W^{\uq} (\bar{\q}_i)$ for the valence quark $\uq$ to join
the antiquark $\bar{\q}_i$ are   
\beq
W^{\uq}(\bar{\uq})=10/23, 
\quad W^{\uq}(\bar{\dq})=10/23,
\quad W^{\uq}(\bar{\sq})=3/23.  
\label{c3}
\eeq
The probabilities $W^{\dq}(\bar{\q}_i)$  that this happens with 
the valence quark $\dq$ are equal to $W^{\uq} (\bar{\q}_i)$.   
Assuming that the fragmentation does not depend on the  probability 
$P_{\q}$  for the  hard interaction, 
the probability to find  a final-state hadron $h_i$ 
after the fragmentation is
\beq
P(h_i)=P_{\q} \> W^{\q} (\bar{\q}_i), \qquad \q=\uq\> \>  \mbox{or}\> \> \dq , 
\qquad \bar{\q}_i= \bar{\dq}, \bar{\uq}, \bar{\sq} ,  
\label{c5}
\eeq
where the flavour of antiquark $\bar{\q}_i$ is 
taken in such a way in order to 
provide a correct flavour content of a meson 
$h_i=(\pi^0,\pi^+,\pi^-,K^0,K^+)$. 
The average charge $\la q^h \ra$  
of the leading current-region  hadrons  is 
\beq
\la q^h \ra = \left[ 1 - P(\pi^0) - P(K^0) \right] 
\sum_{i} q_i^h P(h_i) \simeq 0.25 ,   
\label{c7}
\eeq
where the first factor represents  the probability  
of observing a charged leading particle  
and $q_i^h$ is the electric charge of the
$i$th charged leading meson.       

It should be noted that result (\ref{c7}) stays without
changes even if one considers high-mass  meson states: 
One could split 
$P(h_i)$ into  probabilities for 
different multiplets, but this splitting does
not affect the electric charge of a specific flavour combination
and $\la q^h \ra\simeq 0.25$
still be valid.  
We  have  also verified  that if one uses the suppression factor 0.1 
for diquark production similar to  \cite{jetset}, the contribution of  
the lowest-mass leading baryons ($p$ and $n$) increases
$\la q^h \ra$ by $4\%$ only.        

%%%%%%%%%%%%%%%%%%%%%%%%%%%%%%%%%%%%%%%%%%%%%%%%%%%%%%%%%%%%%%%%% 
\subsection{Contribution of sea quarks and LO QCD effects}
%%%%%%%%%%%%%%%%%%%%%%%%%%%%%%%%%%%%%%%%%%%%%%%%%%%%%%%%%%%%%%%%%%

The interactions with  sea quarks  change (\ref{c7}).
If one considers  the QPM type of  the hard interactions with sea
quarks only, 
the average charge seen in the current region is zero, since quarks and
antiquarks can be knocked out of the 
proton  with equal probabilities. Let us
define $R_{\mathrm{sea}}$ as the probability of the interaction
with sea quarks. Such interactions decrease 
the average  charge by a  factor $(1 - R_{\mathrm{sea}})$,
which is just the probability that the  struck parton happens to be 
one of the valence quarks. 

The first-order QCD processes further decrease (\ref{c7}).
For the BGF events with  two hard partons  moving the target region and  
for events with a single quark
in  the  current region, one obtains $\la q^h \ra =0$
(quark  and antiquark can be emitted to the current region 
with equal probabilities).  Therefore, this further 
decreases $\la q^h \ra$ by a factor $(1 - R_{\mathrm{BGF}})$, 
where $R_{\mathrm{BGF}}$ is the probability for the BGF event to occur.

It is more complicated to include  the QCDC process. 
Let us define  the production rate of the QCDC events as $R_{\mathrm{QCDC}}$.
If only a single quark
moves to the current region, this gives a similar contribution
to the leading particle charge as in the QPM.   However, one can notice
that there is a fraction of QCDC events which does  
not contribute to the average  charge. These events 
can be classified as: 
1) Events with a single hard gluon moving to the current region;
2) Events without partons in the current region;
3) Events with both quark and gluon moving to the current region, in
which the transverse motion of the string connecting both current-region
partons cannot produce a preferable  longitudinal momentum
of hadrons steaming  from the hadronisation of the hard parton.  
We define  a  fraction of the events for these three types  as 
$\tilde{R}_{\mathrm{QCDC}}$ 
($\tilde{R}_{\mathrm{QCDC}} < R_{\mathrm{QCDC}})$.

Taking into account all of the contributions discussed above, one obtains
\beq
\la q^h \ra = 0.25 \> (1 - R_{\mathrm{sea}}) 
(1 - R_{\mathrm{BGF}}) (1 - \tilde{R}_{\mathrm{QCDC}}). 
\label{c10}
\eeq
At a fixed $Q^2$, the values of $R_{\mathrm{sea}}$ and $R_{\mathrm{BGF}}$
decrease with increase of $x$. Thus $\la q^h \ra$ rises as $x$ increases.

From the  point of view of non-perturbative physics,
it is  interesting to investigate the factor
$0.25 \> (1 - R_{\mathrm{sea}})$ in (\ref{c10}) which  
comes from our consideration along the line of
independent or the LUND fragmentation.  Therefore, the cluster
hadronisation  may  give different results.
The contribution
$R_{\mathrm{sea}}$ is interesting  since it contains information
on the proton structure function.

Note that high-order QCD,
hadronisation and resonance decays should further  decrease the $\la q^h \ra$
since they produce an additional  smearing effect which
can properly be taken into account using a MC simulation.

%%%%%%%%%%%%%%%
\subsection{MC Study}
%%%%%%%%%%%%%%%

\begin{figure}[htb]
\begin{center}
\vspace{-2.0cm}
\mbox{\epsfig{file=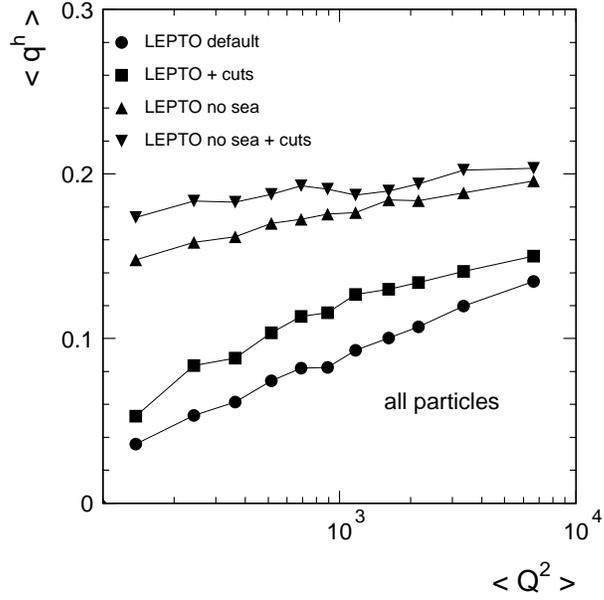,height=10.5cm}}

\vspace{-2.0cm}
\mbox{\epsfig{file=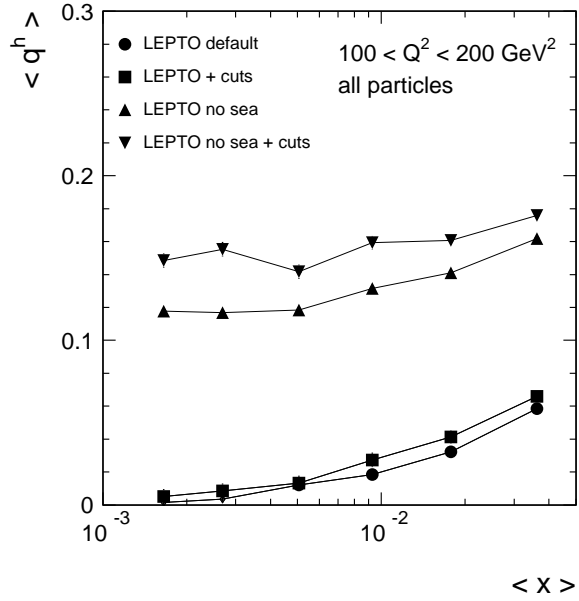,height=10.5cm}}
\end{center}
\caption{{\it Average charge
of leading particles before and after the BGF reduction (the label ``+cuts'').
The leading particles were  determined using  charged and neutral
final-state hadrons generated with LEPTO MEPS. We also show the LEPTO
predictions 
without sea quarks, as described in the text. 
The statistical uncertainties
are smaller than the size of the symbols. }}
\label{fig7}
\end{figure}

Fig.~\ref{fig7} shows the average charge of leading 
particles in LEPTO as a function of $\la Q^2\ra$ and $\la x\ra$.   
A leading particle with a minimum value of $p_Z^{\mathrm{Breit}}$
was identified among  all charged and  neutral 
final-state hadrons  in the current region of the  
Breit frame. 
To investigate the contribution from sea quarks, we generated  
a DIS sample  after the rejection of struck antiquarks 
$\bar{\uq}$ and $\bar{\dq}$ from sea as well as
all heavy flavour sea quarks and antiquarks. 
This can easily  be performed using the LST(25) parameter 
in LEPTO, without redefinition of the structure function. 
Note that this  method does not completely remove  the contribution of 
sea quarks $\uq$ and $\dq$ to  valence quarks in the proton, 
since  LEPTO does not distinguish 
between light valence  and sea quarks \cite{ing}. 
Fig.~\ref{fig7} also  shows the average charge after the use of   
the cuts to reject the BGF contribution. 

According to the simulation, the expected limit $0.25$ is still
not reached for the given energy scale. 
The contribution of sea
quarks is very large and depends on $x$:  At
low $x$, sea quarks dominate and 
this reduces $\la q^h\ra$. The cuts described
in Sect.~\ref{sec:red}
increase the value of the average leading-particle charge, 
as one expects from a decrease of $R_{\mathrm{BGF}}$ in (\ref{c10}).

%%%%%%%%%%%%%%%
\section{Charm}
%%%%%%%%%%%%%%%

The study of the charm  production in DIS is of importance for
the understanding of the sea parton densities in the nucleon. 
The three diagrams shown in Fig.~\ref{fig1} can contribute to the
charm cross  section. There is an evidence that the main fraction
of charm quarks is created in the BGF \cite{H1c}, rather than
through the QPM type of scattering.  This, however, cannot 
be considered as a solid  conclusion yet due to
small statistics available at that time 
and since the LEPTO used to measure $(2\mid \vec{p}_D\mid )/W$ in \cite{H1c} 
had no QCDC process, which is an additional non-BGF contribution 
to the QPM picture at medium $Q^2$. 

In addition to the perturbative QCD charm mechanisms, 
it was suggested long time ago
the hypothesis of intrinsic charm \cite{Brod}. 
Possibilities of probing the intrinsic charm at HERA have been
discussed in \cite{CIng}. 
According to this
model, the proton wave function has an additional
contribution from $\mid uudc\bar{c} \ra$. 
In this model the intrinsic charm is  
created  due to scattering either on the intrinsic charm quark or
a light valence quark, so that intrinsic $c\bar{c}$ pair moves 
along the proton remnants. For these cases, 
the only DIS  processes contributing  to the production of 
intrinsic  charm are the QPM and  QCDC. 

One of the important issues in the intrinsic charm studies
is  to find a  method  to isolate the QPM and QCDC processes from the BGF.
Having obtained a subsample with a small  
fraction of the BGF type of DIS events, one can perform  a more detailed  
study of charm. Some ideas on  how  to suppress the  BGF background  
in the  intrinsic  charm study has already been discussed in \cite{CIng}. 

To illustrate  the  possibility to reduce the events with charm 
coming from the BGF using our method,  
we generated  the charmed DIS events with
AROMA 2.1 \cite{aroma}  and LEPTO 6.5 MEPS 
models.  
We used the CTEQ4M \cite{pdf} structure
function for both simulations.
The AROMA models the charm production
through the BGF, which is implemented at the  LO including heavy quark
mass.  
In the LEPTO, the BGF processes were turned off, so that
all charm quarks originate  from  sea in the proton 
(charm from fragmentation can be safely neglected). 
The intrinsic  charm was not
included explicitly in this simulation: The kinematic signatures of the
intrinsic charm are expected to be similar 
to those of the sea quark production in QPM/QCDC \cite{CIng}.     

Applying the selection described in Sect.~\ref{sec:red},
only about $2-3\%$ of the BGF events with charm  passed the cuts.
On the other hand, more than $20\%$ of the events with sea charm  
survive the cuts, i.e. the suppression of BGF events  
with charm  is  by a factor 8-10 larger than for the QPM/QCDC type
of charm production. 

The  reconstructed 4-momenta of charm candidates, transformed  
into the Breit frame, can also be used to distinguish between 
QPM/QCDC and BGF type of the LO events. This can be done by
measuring the transverse current-region momentum of a charm hadron.
Such  a transverse momentum     
is expected to be softer for QPM/QCDC than for BGF. 
Another possible approach is to calculate the ratio 
$R=\la N_{\mathrm{target}}\ra /\la N_{\mathrm{current}}\ra$,
where $\la N_{\mathrm{target}}\ra$ ($\la N_{\mathrm{current}}\ra$) is
the average multiplicity of charm hadrons in the target (current)
region. For QPM, $R=0$, while for the BGF, $R>1$. The QCDC gives also
$R\ne 0$, but this value is not as large as for the BGF 
(see Fig~\ref{fig4}).  After applying  cuts 
on the ratio $R$ or the transverse
current-region  momenta of charm candidates,
one could obtain a sample of non-BGF charm events to be studied further.      
  
After the BGF suppression,  
the resulting DIS events might contain
a fraction of events with intrinsic charm,  large enough  to
obtain an experimentally detectable excess in the charm production 
over the conventional QCD  processes. This  excess can be seen at
not very high $Q^2$ and  a sufficiently large $x$, i.e. in 
the phase-space regions where the BGF has  
an additional  suppression.

%%%%%%%%%%%%%%%%%%%%%%%%%%%%%%%%%%%%%
\subsection{Conclusion}
%%%%%%%%%%%%%%%%%%%%%%%%%%%%%%%%%%%%%

In this paper we show that, in the LO  formalism with
multiple parton emission described by a  parton shower,  
one can obtain a subsample of DIS events  
where the probability to observe QPM/QCDC type of events 
with (anti)quark recoiled against the proton 
is clearly enhanced,  in contrast
to DIS events without any preselection.  
This can be done without jet clustering
algorithms that reject both QCDC and BGF type of events
and suffer from  ambiguity in jet definitions and 
misclustering at low  $Q^2$ (jet $E_T$).  
The investigation of this subsample can provide a deeper insight
into the proton structure  by studying the
flavour of the struck quark, indirectly, through the  measurement of the
leading particle charge or, more directly, by reconstructing  the
heavy flavour quarks. More explicitly, 
the first measurement is very sensitive
to the sea quark contribution to the proton structure function 
and details in the
fragmentation mechanism.  The second study may  allow
to observe an excess in charm production over the
conventional perturbative QCD  mechanisms and thus    
to set an upper limit on the intrinsic charm 
inside the proton in DIS at HERA.       

Note that the 
efficiency of the BGF rejection can be    
different for different cut-offs on the matrix elements.
To study this, it is necessary to re-tune the LEPTO model  with a new
values of cut-offs in order to obtain    
a good description of DIS data. 
Moreover, rather significant  
reduction of  the BGF processes which seen in LEPTO 
might be a feature of any  Monte Carlo model  based on  
matching of the first-order QCD matrix elements  
with  softer emissions in  parton showers. 
Presently, it is difficult to verify  this point since
less formal models based solely on parton showers, ARIADNE and HERWIG,   
do not distinguish between different types of the LO processes.   

\section{Acknowledgments:}
I thank L.~Gladilin, S.~Magill and G.~Ingelman 
for valuable  discussions.

{}

\newpage 
%%%%%%%%%%%%%%%%%%%%%%%%%%%%%%
%%%%%%%%%%%%%%%%%%%%%%%%%%%%%%

%%%%%%%%%%%%%%%%%%%%%%%%%%%%%%
%%%%%%%%%%%%%%%%%%%%%%%%%%%%%%%

\newpage 
%%%%%%%%%%%%%%%%%%%%%%%%%%%%%%
%%%%%%%%%%%%%%%%%%%%%%%%%%%%%%%

\newpage
%%%%%%%%%%%%%%%%%%%%%%%%%%%%%%
%%%%%%%%%%%%%%%%%%%%%%%%%%%%%%%

\newpage 
%%%%%%%%%%%%%%%%%%%%%%%%%%%%%%
%%%%%%%%%%%%%%%%%%%%%%%%%%%%%%%

\newpage 
%%%%%%%%%%%%%%%%%%%%%%%%%%%%%%
%%%%%%%%%%%%%%%%%%%%%%%%%%%%%%%

\newpage 
%%%%%%%%%%%%%%%%%%%%%%%%%%%%%%
\begin{figure}[htb]
\begin{center}
\vspace{1.0cm}
\mbox{\epsfig{file=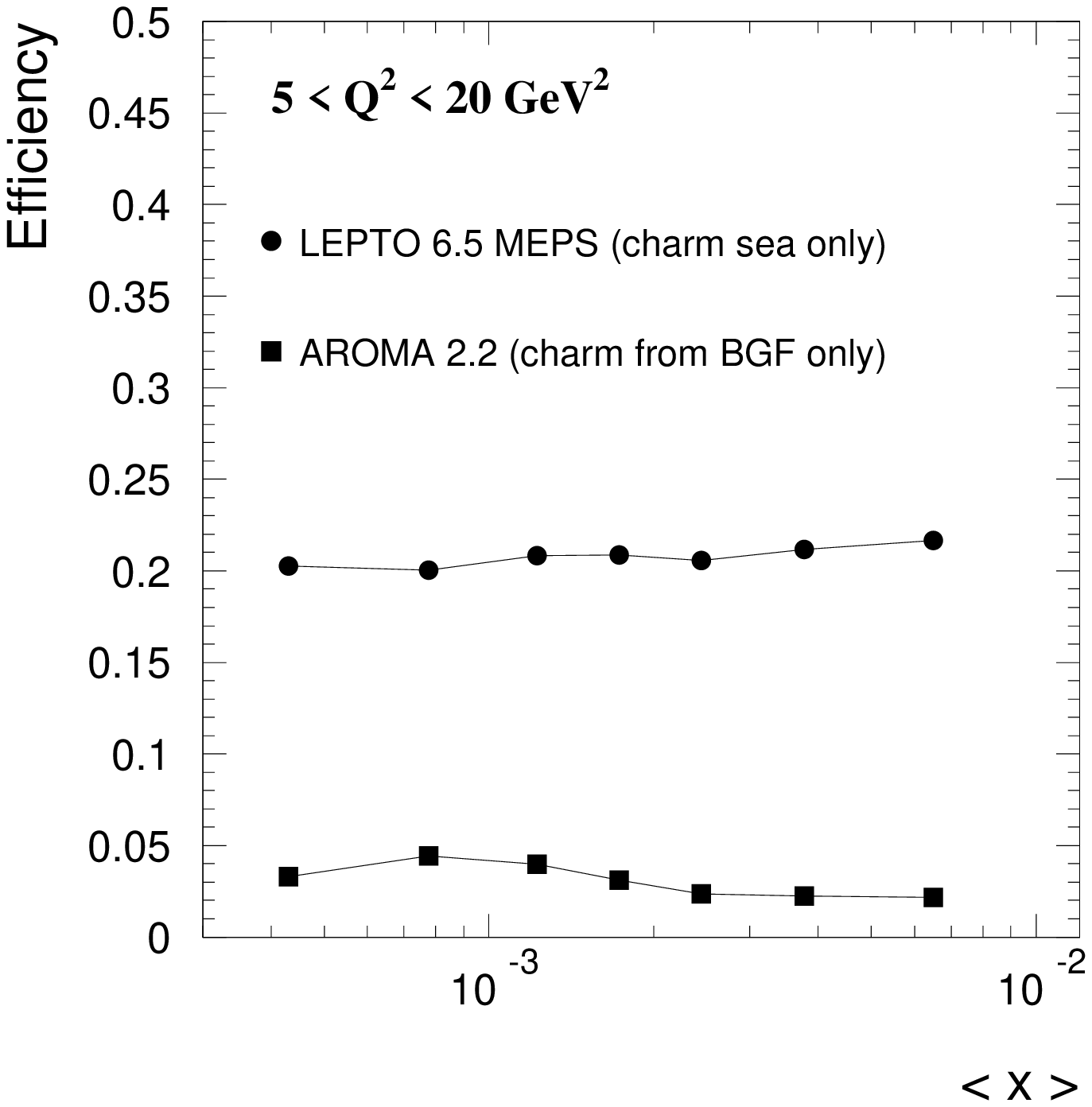,height=12.0cm}}
\end{center}

\vspace{-1.5cm}
\caption{{\it
The efficiency of the event selection after the BGF reduction for
AROMA (charm comes from only BGF) and LEPTO (charm is created in
the QPM/QCDC processes) Monte Carlo models.}}
\label{fig9}
\end{figure}
%%%%%%%%%%%%%%%%%%%%%%%%%%%%%%%


\newpage         
\begin{thebibliography}{99}

\bibitem{MGHERA1}
T.~Carli, J.~Gerigk, A.~Ringwald, F.~Schrempp, these proceedings 

\bibitem{MGHERA2}
B.~Delcourt, these proceedings 

\bibitem{MGHERA3}
A.~Edin, G.~Ingelman, these proceedings    

\bibitem{Br}
R.~P.~Feynman, Photon-Hadron Interactions, Benjamin, NY, 1972

\bibitem{chek}
S.~V.~Chekanov,  \Journal{\JPG}{25}{59}{1999}; 
S.~V.~Chekanov, Presented at XXVIII International Symposium
on Multiparticle Dynamics, 1998, (Delphi, Greece)
hep-ph/9810477

\bibitem{LEP}
G.~Ingelman, A.~Edin and  J.~Rathsman,
\Journal{\CPC}{101}{108}{1997}

\bibitem{jetset}
T.~Sj\"ostrand, \Journal{\CPC}{82}{74}{1994}

\bibitem{ARD}
L.~L\"{o}nnblad,  \Journal{\CPC}{71}{15}{1992}
 
\bibitem{HEW}
G.~Marchesini et al.,  \Journal{\CPC}{67}{465}{1992}

\bibitem{jur}
J.~G.~K\"{o}rner, E.~Mirkes, G.~A.~Schuler, 
\Journal{\IJMPA}{4}{1781}{1989} 

\bibitem{str}
K.~H.~Streng, T.~F.~Walsh and  P.~M.~Zerwas,   
\Journal{\ZPC}{2}{237}{1979}

\bibitem{GRV}
M.~Gl\"uck, E.~Reya and A.~Vogt,
\Journal{\PLB}{306}{391}{1993} 

\bibitem{pdf}
H.~Plothow-Besch, {\it PDFLIB: Nucleon, Pion and Photon Parton
Density Functions and $\alpha_s$ Calculations} (1997)  
User's Manual CERN-PPE 

\bibitem{ing}
G.~Ingelman, privite communication 

\bibitem{H1c}
H1 Coll., C.~Adloff et al., \Journal{\ZPC}{72}{593}{1996} 

\bibitem{Brod}
S.~J.~Brodsky et al., \Journal{\PLB}{93}{451}{1980}; 
S.~J.~Brodsky and C.~Peterson, \Journal{\PRD}{23}{2745}{1981} 

\bibitem{CIng}
G.~Ingelman, J.~J\"{o}nsson and M.~Nyberg, \Journal{\PRD}{47}{4872}{1993}; 
G.~Ingelman, M.~Thunman, \Journal{\ZPC}{73}{505}{1997}    


\bibitem{aroma}
G.~Ingelman, J.~Rathsman and G.~A.~Schuler,
\Journal{\CPC}{101}{135}{1997}  

\end{thebibliography}
\end{document}